# The Brightening of Saturn's F Ring

Robert S. French[a], Mark R. Showalter[b], Rafael Sfair[c], Carlos A. Argüelles[d], Myriam Pajuelo[e], Patricio Becerra[f], Matthew M. Hedman[g], and Philip D. Nicholson[h]

[a] SETI Institute, 189 Bernardo Ave., Mountain View, California 94043, United States; rfrench@seti.org
[b] SETI Institute, 189 Bernardo Ave., Mountain View, California 94043, United States; mshowalter@seti.org
[c] UNESP – Sao Paulo State University, Guaratinguetá, CEP 12.516-410 SP, Brazil; rsfair@feg.unesp.br
[d] Sección Física, Departamento de Ciencias, Pontificia Universidad Católica del Perú, Apartado 1761, Lima, Perú; c.arguelles@pucp.edu.pe
[e] Sección Física, Departamento de Ciencias, Pontificia Universidad Católica del Perú, Apartado 1761, Lima, Perú; mpajuelo@pucp.edu.pe
[f] Comisión Nacional De Investigación y Desarrollo Aeroespacial (CONIDA), Ca. Luis Felipe Villarán 1069, San Isidro, Lima, Peru; becerra@lpl.arizona.edu; Present address: Lunar and Planetary Laboratory, University of Arizona, 1629 E. University Blvd., Tucson, Arizona 85721, United States
[g] Astronomy Dept., Space Sciences Building, Cornell University, Ithaca, New York 14853, United States; mmhedman@astro.cornell.edu
[h] Astronomy Dept., Space Sciences Building, Cornell University, Ithaca, New York 14853, United States; nicholso@astro.cornell.edu

## ABSTRACT

Image photometry reveals that the F ring is approximately twice as bright during the Cassini tour as it was during the Voyager flybys of 1980 and 1981. It is also three times as wide and has a higher integrated optical depth. We have performed photometric measurements of more than 4,800 images of Saturn's F ring taken over a five-year period with Cassini's Narrow Angle Camera. We show that the ring is not optically thin in many observing geometries and apply a photometric model based on single-scattering in the presence of shadowing and obscuration, deriving a mean effective optical depth $\tau \approx 0.033$. Stellar occultation data from Voyager PPS and Cassini VIMS validate both the optical depth and the width measurements. In contrast to this decades-scale change, the baseline properties of the F ring have not changed significantly from 2004 to 2009. However, we have investigated one major, bright feature that appeared in the ring in late 2006. This transient feature increased the ring's overall mean brightness by 84% and decayed with a half-life of 91 days.

# 1. INTRODUCTION

Ever since its discovery by the Imaging Photopolarimeter on Pioneer 11 (Gehrels et al., 1980), Saturn's F ring has been a source of continuing surprises. The two Voyager spacecraft provided the first detailed images of the F ring, revealing a narrow, relatively dense central core and a lower-density surrounding skirt. The ring showed a high degree of longitudinal variation including the well-known kinks, clumps, strands, and so-called braids (Smith et al., 1981, 1982); the brightness of the ring varied by a factor of more than three from one longitude to another. Observations from the Hubble Space Telescope in 1995 showed similar variations in brightness (Nicholson et al., 1996). More recently, the instruments on the Cassini spacecraft have provided thousands of images and measurements of the F ring in unprecedented detail. They continue to show dramatic variations in azimuth and time (Charnoz et al., 2005; Murray et al., 2005; Porco et al., 2005).

Keplerian shear should smooth out longitudinal variations in a few hundred orbits, so their continuing presence implies a highly dynamic environment where the variations are continually regenerated. It is now clear that many of the features are related to the ring's gravitational interactions with Prometheus (Murray et al., 2008). Further, there is evidence for embedded moonlets that inject fine material into the F ring as the result of collisions in the F ring's core (Nicholson et al., 2007; Esposito et al., 2008; Murray et al., 2008; Charnoz et al., 2009) or perhaps after being impacted by small meteoroids (Showalter, 1998).

Showalter et al. (1992) analyzed the F ring using photometry from Voyager 1 and 2 cameras. They measured 214 radial profiles from 69 images at a variety of phase angles, and the results were then fit to particle models (Pollack and Cuzzi, 1980) to determine the particle properties and size distribution. The model fit implied that basically all scattering of light from the F ring is due to an optically thin ensemble of microscopic dust particles. However, the large longitudinal variations and the relatively small number of radial profiles available from Voyager reduced the precision with which photometric analysis could be done.

In this study, we expand upon the work of Showalter et al. (1992) by performing a more extensive analysis using Cassini images. The primary improvement provided by our study comes from the large quantity and high quality of data now available and from the new image processing tools available. We have also reprocessed the original images used by Showalter et al. (1992) using modern calibration and processing techniques, resulting in more accurate photometric measurements from that data set. The availability of two data sets provides a unique opportunity to compare the gross properties of the F ring over a time span of 30 years. We further augment this photometric analysis with stellar occultation data.

In Section 2 we describe the selection and processing of the Cassini and Voyager images. In Section 3 we discuss the photometric measurements from the images, which reveal significant changes between the two data sets. In Section 4 we analyze stellar occultation profiles from the Voyager PPS and Cassini VIMS instruments, which show associated changes. In Section 5 we explore variations in the ring's radial width. In Section 6 we discuss instrument considerations. Finally, in Section 7 we discuss our results and their implications.

## 2. IMAGE PROCESSING

*2.1. Cassini image selection and processing*

The Cassini Imaging Science Subsystem (ISS) consists of two cameras: the Narrow Angle Camera (NAC), with a 0.35° × 0.35° field of view (FOV) and the Wide Angle Camera (WAC) with a 3.5° × 3.5° FOV (Porco et al., 2004). Due to its much higher spatial resolution, we use NAC images exclusively in this study. Unless otherwise specified, we use images obtained through the clear filters (CL1 and CL2) for all photometry. The total system transmission using the clear filters has an effective wavelength $\lambda_{eff}$ = 651 nm using the solar spectrum; it has > 50% sensitivity from 500 to 850 nm, with moderate sensitivity from 200 to 500 nm and significantly decreased sensitivity beyond 850 nm.

During the Cassini tour, the NAC has taken a series of "movies" of the F ring. Each movie contains from a few dozen to over 1,000 images taken over a period of 10 to 15 hours. By staring at a nearly fixed longitude, most often the ring ansa, Cassini obtains nearly complete longitudinal coverage of the F ring as it rotates under the spacecraft. We base our analysis entirely on these movies. Twenty-five of these movies, totaling 4,810 images, were examined for this paper (Table 2, end of paper). All images were retrieved from the Rings Node of NASA's Planetary Data System.

The Cassini images were calibrated using the "CISSCAL" pipeline (2004 version). This pipeline includes uneven pixel gain compensation, bias subtraction, dark frame subtraction, non-linear gain compensation, flat-field correction, and adjustment for optical distortions and filter passbands. The resulting image pixels are calibrated as *I/F*, where intensity *I* is measured relative to the incident solar flux density $\pi F$ and scaled such that *I/F* = 1 for a perfectly diffusing, flat Lambertian surface illuminated from normal incidence. Because our images are of the ring ansa and the F ring predominantly forward-scatters, we do not expect Saturnshine to have a noticeable influence on our results (Showalter et al., 1992).

Each image was processed to produce a single radial profile of the F ring as close to the ring ansa as possible (Fig. 1). The use of the ansa provided the finest radial resolution and also simplified image processing. Due to the large number of images and to the automatic longitudinal averaging provided by the movies, it was unnecessary to take multiple radial measurements within each image. However, we used a "thick" radial profile that averaged together many pixels within each image perpendicular to the radial direction. We found that the measurement was generally insensitive to the thickness chosen, with resulting values varying by at most 3%. As a result, the exact location of each radial profile and its width were simply chosen by eye to avoid obvious image contaminants such as bad pixels, missing data, or the presence of Prometheus or Pandora, while still allowing for moderate local averaging. Whenever possible, the same radial position and thickness were used for each frame of a movie sequence, with minor corrections made for changes in Cassini camera pointing. As the images were usually taken at fixed intervals, this provided nearly uniform sampling of the ring. Although some rotation movies provided less than 360° of coverage (and a few contained more), we could confirm by examining adjacent ring movies, which usually covered the missing areas, that there were no significant features in the missing regions that would have biased our samples.

Each resulting profile was integrated to produce the *equivalent width*, defined as:

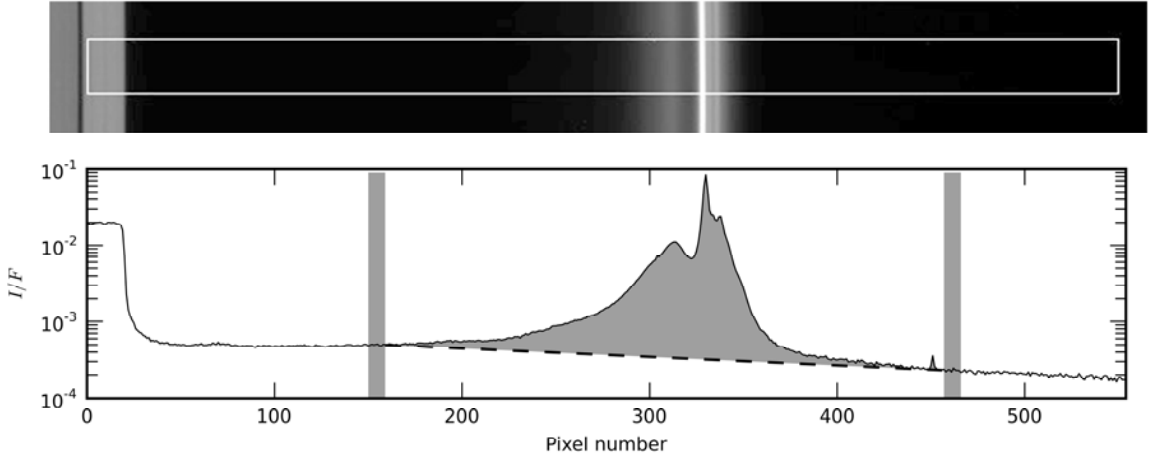

Fig. 1: An example of measuring the radial profile of the F ring. At the top is a section of image N1545556618_1 ($\alpha = 158.58°$, $e = 125.73°$, $i = 75.5°$). The image has been non-linearly stretched to make the environment of the F ring more obvious. A portion of the A ring (and the Keeler gap) is shown on the left and the F ring is on the right. The white bounding box was used to define the dimensions for the radial profile. Pixels were averaged in the vertical (longitudinal) direction to produce the measurements shown. The vertical gray rectangles show the portion of the profile that was used to determine the background values inside and outside of the F ring, and the dashed line shows the linearly interpolated background that was subtracted from the *I/F* measurements. Finally, the shaded area under the curve shows the area that was integrated to create the equivalent width (*W*).

$$W = \int I/F(a)da \qquad (1)$$

where *a* is the radial distance from Saturn. (Note that, by convention, we treat "*I/F*" as a syntactic unit throughout this paper, so the radial brightness profile is written as *I/F(a)* rather than as the more formally correct *I(a)/F*). The equivalent width is the radial width of a perfectly reflecting flat Lambertian surface, illuminated from normal incidence, that would provide the same total intensity as that measured from the ring. The use of equivalent width allows the comparison of images with different spatial scales.

We have employed an automatic procedure to locate the outer edge of the A ring, which served as the geometric reference for our ring profiles. The photometric center of the F ring was detected automatically, and the halfway point between the edge of the A ring and the F ring was defined as the inner limit of our integrations. The same distance on the opposite side of the F ring defined the outer limit. This resulted in a radial extent for the integration of approximately 138,465 km to 141,835 km, a width of 3,370 km. Visual examination of the light curves showed that this algorithm easily enclosed the F ring in all cases. A representative set of pixels (usually nine, assuming the gap between the A ring and the F ring was wide enough) at the inner and outer edge of the integration limits was used to determine a background value to account for any remaining background bias. This bias was assumed to vary linearly from the inner edge to the outer edge and was subtracted from the radial profile before further processing. Although dust has been detected interior to the F ring that could bias the background subtraction, the dust has been shown to have an optical depth of ~$10^{-4}$ (Showalter et al., 1998) making any contribution negligible for our purposes. Figure 1 shows an example of this procedure.

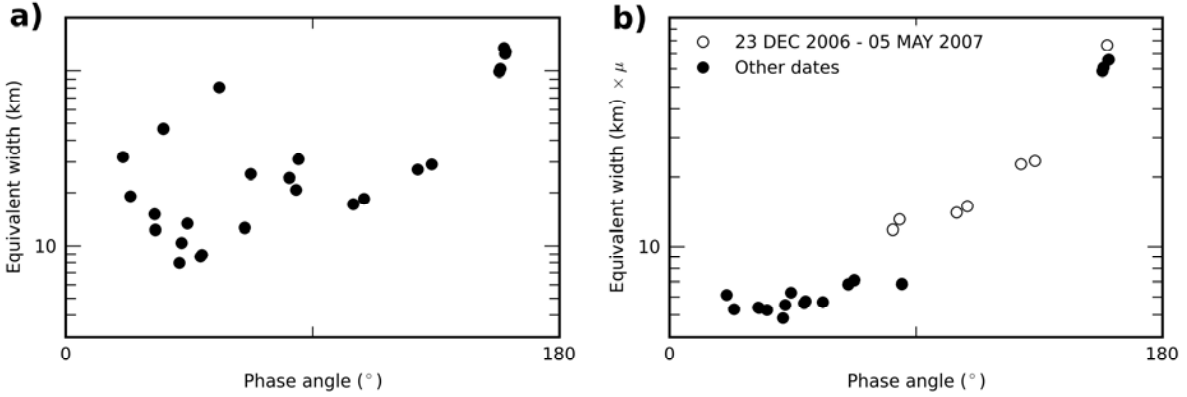

**Fig. 2: Cassini NAC movie measurements. a)** The mean equivalent width ($W$) for all Cassini NAC movies listed in Table 2. Error bars are not shown because the variation of the mean is only 2–3%. **b)** Equivalent width adjusted for emission angle ($W_\perp$). Observations taken immediately after the formation of the clump in late 2006 are indicated as open circles. The $W_\perp$ values of these images is systematically larger than those of observations taken at other dates but similar phase angles.

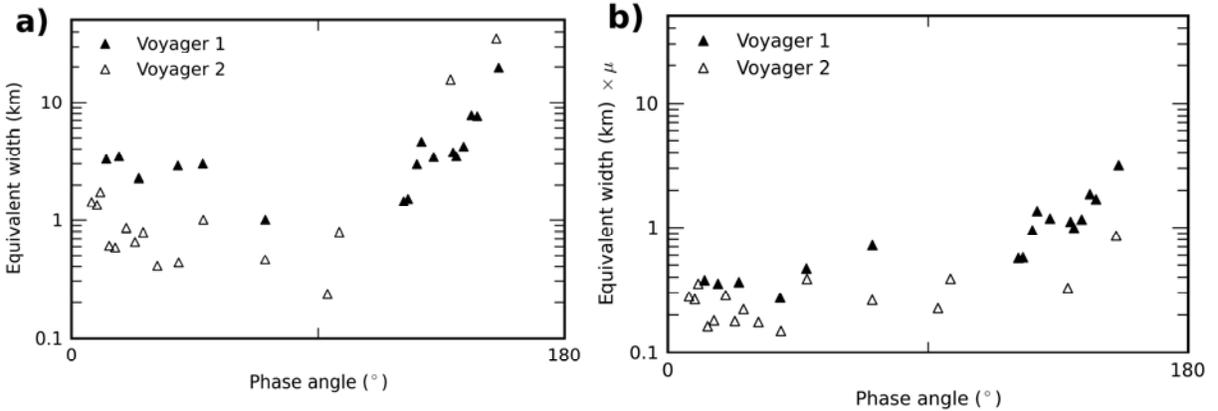

**Fig. 3: Voyager measurements. a)** The equivalent width ($W$) for all binned Voyager measurements. **b)** Equivalent width adjusted for emission angle ($W_\perp$).

For the purposes of this study, measurements (one per image) were averaged together to produce a single result per movie. Despite the longitudinal variability intrinsic in the F ring, the resulting standard errors of the means were on the order of 2–3%. The phase, emission, and incidence angles, which did not vary significantly during the time required to obtain each movie, were also averaged. The resulting equivalent widths are shown in Fig. 2a as a function of phase angle $\alpha$, the angle between the direction to the Sun and the direction to the observer as measured at the ring.

In the F ring, we already expect $W$ to be highly variable as a function of longitude and time. We wish to decouple these variations from those defined purely by the viewing and lighting geometry. A first-order correction can be applied by assuming that the ring's optical depth $\tau$ is small. In this limit, following the work of Showalter et al. (1992), all particles are uniformly illuminated by the Sun so the dependence on the incidence angle $i$ disappears. The effect of viewing geometry is to make a ring's surface reflectivity proportional to $1/|\cos(e)|$, where $e$ is the emission angle, measured from the ring plane normal vector on the lit side of the ring to the direction to the observer. Using $\mu \equiv |\cos(e)|$, we define the "normal equivalent width" $W_\perp$ as:

$$W_\perp = \int \mu I/F(a)da \equiv \int I/F_\perp(a)da \qquad (2)$$

Because $\mu$ is approximately constant for a given observation, it can be moved outside the integral leaving $W_\perp = \mu W$. Figure 2b shows the Cassini measurements with this adjustment applied; note that the scatter among the measurements at similar phase angles has been substantially reduced. The remaining variations illustrate the ring's phase curve, which brightens markedly toward high $\alpha$ due to diffraction by the ring's population of fine dust. Showalter et al. (1992) found this correction by a factor of $\mu$ to be sufficient for their analysis of Voyager data. However, the finer precision of our Cassini measurements impels us to apply additional refinements; these are discussed below in Section 3.

*2.2. Voyager image selection and processing*

Like Cassini, the Voyager 1 and 2 Imaging Science Subsystem (Smith et al., 1977) consists of narrow- and wide-angle cameras. The spectral response of their selenium-sulfur vidicon detectors is significantly narrower than that of Cassini's CCDs, with $\lambda_{eff}$ = 497 nm and a > 50% sensitivity of 375 to 525 nm for the Voyager NAC and $\lambda_{eff}$ = 470 nm and a > 50% sensitivity of 400 to 575 nm for the Voyager WAC.

Out of computational necessity, Showalter et al. (1992) had employed raw images in their original analysis. The same images used by Showalter et al. were reprocessed for this study using the complete Voyager pipeline for calibration and geometric correction. Analysis techniques were otherwise similar to those described above, except that longitudinal coverage was very limited, many images had low signal-to-noise ratio, and some images had insufficient resolution to separate the A and F rings cleanly. The equivalent widths were computed as described above. Measurements from images with similar phase, incidence, and emission angles were combined to improve photometric precision. The resulting measurements of $W$ and $W_\perp$ are shown in Fig. 3.

## 3. IMAGE PHOTOMETRY

A number of additional matters need to be considered before we can draw robust conclusions from the essentially raw data of Fig. 2. We seek the "baseline" properties of the ring, after various transients, artifacts, and viewing dependencies have been removed.

*3.1. The bright feature of late 2006*

A major, bright feature suddenly appeared in the F ring near the end of 2006 (Fig. 4). Although similar to other small variations that are common in the F ring, this feature is unusual in its brightness, longitudinal extent, and long lifetime. Its effect on the mean photometry of the F ring can be seen in Fig. 2b, where measurements taken during the December, 2006 to May, 2007 period are systematically brighter, for a given phase angle, than measurements taken at other times.

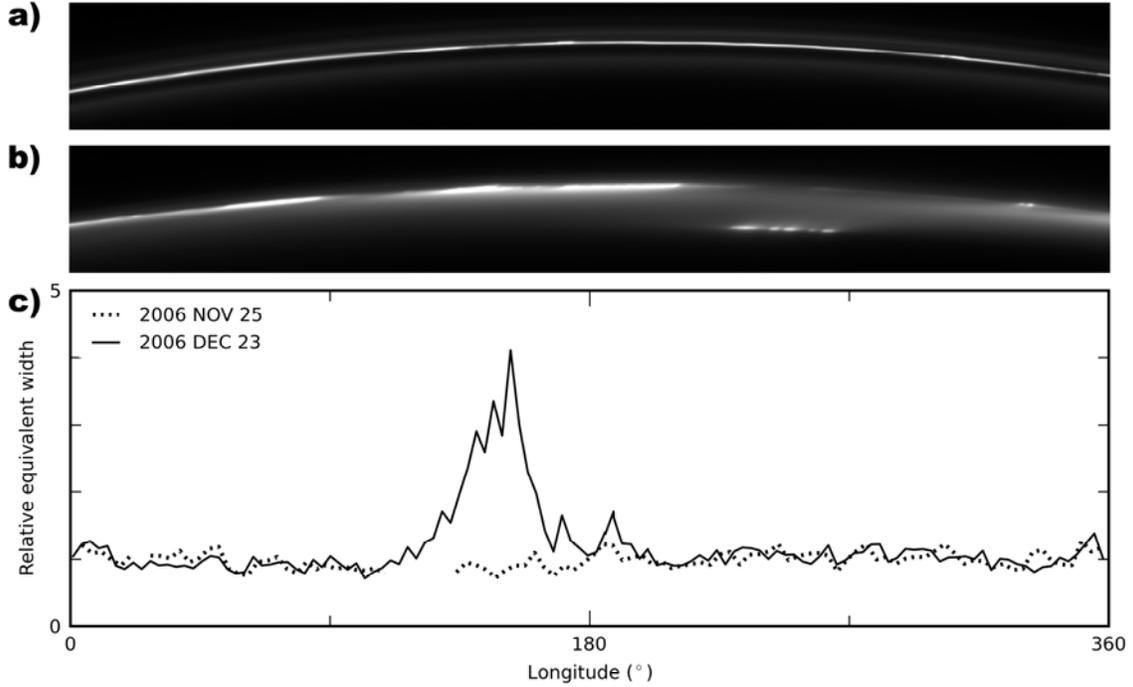

**Fig. 4: The feature of late 2006. a) A 9° portion of the F ring (longitudes 146.3–155.4) as seen on November 25, 2006 (image N1543214356). b) A similar 9° portion (longitudes 148.1–157.1) as seen on December 23, 2006 (image N1545564196, corresponding to JPL Photojournal image PIA08863). c) The longitudinal profile of the movies from observation IDs ISS_033RF_FMOVIE001_VIMS taken on November 25, 2006 (dotted) and ISS_036RF_FMOVIE001_VIMS (solid) taken on December 23, 2006.**

To study the ring's baseline properties, we need to compensate for this unusual feature, which initially contributes more than three times as much light as the rest of the ring combined. Later profiles show the feature expanding in longitude and diminishing in brightness, such that after two years the ring had nearly returned to its prior state.

We have modeled the contribution of this feature as linear brightening followed by an exponential decay:

$$W_{\perp,corrected} = \begin{cases} W_{\perp,measured} & t < t_0 \\ W_{\perp,measured}\left(1 + \dfrac{h(t-t_0)}{t_1-t_0}\right) & t_0 \leq t < t_1 \\ W_{\perp,measured} / (1 + h e^{-c(t-t_1)}) & t \geq t_1 \end{cases} \qquad (3)$$

where $t$ is the time of the observation, $t_0$ is the time that the clump first appeared, $t_1$ is the time at which the clump began to decrease in brightness, $h$ is the peak brightness increase due to the feature at $t_1$, and $c$ is the decay constant.

Because the phase angle varies from one image set to the next, we must simultaneously fit both a decay model and a phase function to the measurements. The ring's phase dependence is described by a phase function $P(\alpha)$. The phase function contains a wealth of information about the sizes and properties of the ring particles (cf. Showalter et al., 1992). However, for our purposes here, we simply model log($P(\alpha)$) via a third-order polynomial; this will be shown to be valid below.

For this model to work, we must assume that the baseline brightness of the ring does not change during the lifetime of the feature. We must also assume that the material in the feature has the same phase function as the ring itself. In addition, we account for the effects of shadowing and obscuration as discussed in the next section. With these assumptions, we can isolate the effects of this feature by using Eq. 3 (and Eq. A.13, derived in the Appendix) to model measurements that lie off the ring's mean phase curve. In a simultaneous fit for nine parameters (times $t_0$ and $t_1$, the peak amplitude $h$, decay constant $c$, the four coefficients of the phase curve, and the effective optical depth discussed below), we determine a peak increase of $h = 84\pm10\%$ starting November 26, 2006 ($\pm12$ days), with the decay beginning April 16, 2007 ($\pm13$ days) and a decay constant of $c = 8.8^{+2.5}_{-1.7}\times10^{-8}\,\mathrm{sec}^{-1}$ (yielding a half-life of $91^{+113}_{-71}$ days). In all cases uncertainties are found using a 70% confidence interval from a chi-squared analysis varying one variable while holding the others constant. The brightness trend is shown in Fig. 5.

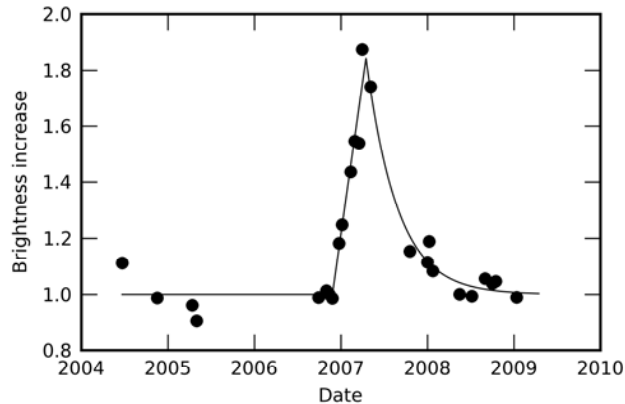

**Fig. 5: Model of the brightening effect of the late 2006 feature. The data points show the Cassini measurements adjusted for $\tau_{equiv}$ = 0.033 and divided by the best-fit phase curve.**

*3.2. Shadowing and obscuration*

As noted above, the reflectivity of an optically thin ring varies in proportion to $1/|\cos(e)|$. Showalter et al. (1992) found this correction alone to be sufficient for their analysis. However, in our more precise analysis, we have noted a residual dependence of $W_\perp$ on the incidence and emission angles. When either is near 90°, so that the Sun or Cassini is close to the ring plane, our measurements of $W_\perp$ are systematically reduced (Fig. 6a). This suggests that the ring is not as optically thin as had been previously supposed.

We present in the Appendix a formalism that models the dependence of $W$ on $i$ and $e$ and accounts for first-order effects of $\tau$ taking into account shadowing and obscuration with single scattering. The end result of our formalism is a new quantity, $\tau_{equiv}$, that represents the mean properties of the ring. In effect, it is the optical depth that would describe a uniform ring with the same mean optical properties as the one we observe. We also produce a new equivalent width, $W_\tau$, that adjusts $W$ for the effects of shadowing and obscuration. We solve for the value of $\tau_{equiv}$ that does the best job of eliminating the dependence of the ring's mean $W_\tau$ on $i$ and $e$. In practice, this is the value that minimizes the residuals when the 25 longitudinally-averaged measurements are fitted to a model where log($P(\alpha)$) is a cubic polynomial in $\alpha$.

Our model actually involves a simultaneous fit for $\tau_{equiv}$, the properties of the December 2006 impact, and the polynomial model of the phase function. We find $\tau_{equiv} = 0.033\pm0.008$ (where the uncertainty is the 70% confidence interval from a chi-squared analysis). The adjusted data are shown in Fig. 6b. Note that the brightness model of Fig. 5 is clearly dependent on the precise equivalent optical depth chosen, and thus our deduced parameters should only be considered approximate. To apply this same technique to the Voyager measurements, we simultaneously used the data from Voyager 1 and Voyager 2 and introduced a scale factor to account for differences in brightness between the two missions. Allowing the phase curve to freely fit the data yielded $\tau_{equiv} = 0.036^{+0.027}_{-0.023}$ with a scale factor of 2.0. Constraining the phase curve to the same shape as the one found for Cassini yielded $\tau_{equiv} = 0.048^{+0.030}_{-0.025}$ with a scale factor of 1.8 and somewhat higher residuals. The $\tau_{equiv}$ for the Cassini measurements and those for Voyager using either technique are consistent within the 70% uncertainties and we will use $\tau_{equiv} = 0.036$ for Voyager in the comparison of the Cassini and Voyager measurements below. The occultation data sets discussed in Section 4 are compatible with both the Cassini and Voyager results presented here.

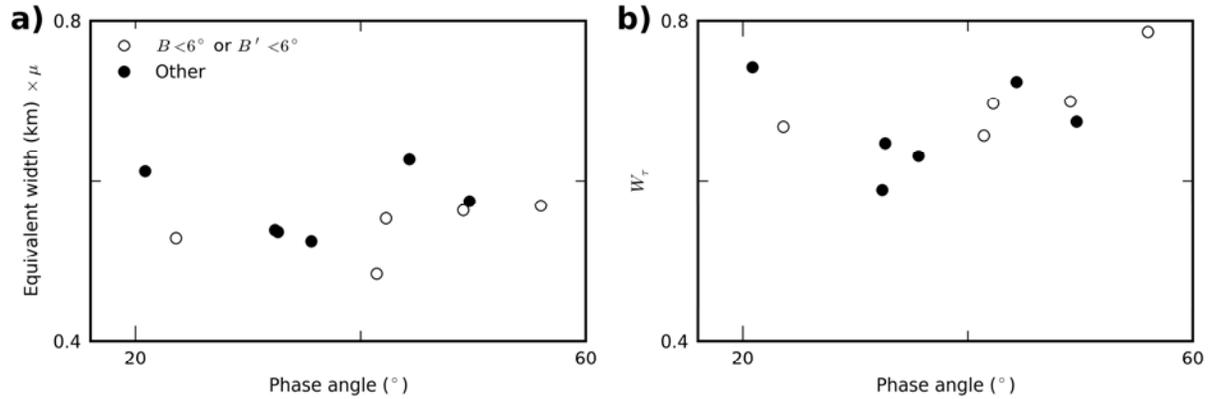

**Fig. 6: Cassini NAC movies adjusted for shadowing and obscuration. a)** The geometrically-corrected equivalent width ($W_\perp$) for low-phase Cassini movies. Observations at low solar elevation angles ($B' = 90\text{-}i$) or low ring opening angles ($B = |90\text{-}e|$) are systematically less bright than other observations at similar phase angles. **b)** The same data adjusted for $\tau_{equiv} = 0.033$.

*3.3. Voyager-Cassini comparison*

At this point we are able to compare the absolute brightness of the F ring during the Voyager and the Cassini missions. We do this by taking the Cassini-derived phase curve, modeled as a cubic polynomial, and scaling it to the Voyager 1 and Voyager 2 measurements. The results are shown in Fig. 7. The mean ratio of the Cassini curve to the Voyager 1 curve is 1.3±0.1 and the mean ratio of the Cassini curve to the Voyager 2 curve is 2.4±0.1, implying that the F ring is substantially brighter at the time of the Cassini observations than it was during the Voyager missions 30 years earlier. The F ring also appears to have decreased markedly in brightness between Voyager 1 and Voyager 2 (mean ratio 1.8±0.2), a result not previously noted by Showalter et al. (1992) but visible in their measurements. Aside from the changes of scale, the shape of the phase curve appears to be generally unchanged. This suggests that although the number of dust particles has been changing, the size distribution of that dust has been relatively stable.

To evaluate the evolution of the brightness of the F ring over the five years of the Cassini data, we have divided the images into subsets by calendar year. We used the same shape of curve described previously and scaled it to the mean of the equivalent width measurements for each year. These curves are plotted in Fig. 8. Each year's mean change from the full curve fit is less than 7%. Once the effects of the December 2006 event have been removed, we see no evidence for systematic changes in the baseline brightness of the F ring between 2004 and 2009.

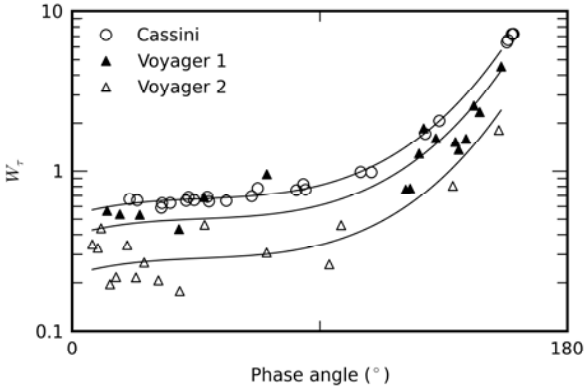 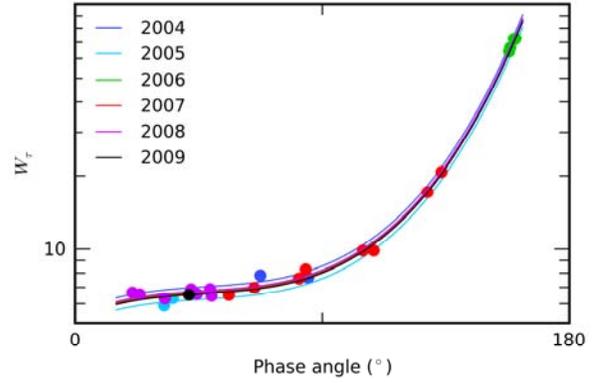

**Fig. 7:** All Cassini and Voyager measurements and their associated 3rd-order fit polynomials. In each case the shape of the fit curve is constrained to be the same as the Cassini curve. The bottom curve is the fit to Voyager 2, the middle curve is the fit to Voyager 1, and the top curve is the fit to Cassini. The mean ratio Cassini-to-Voyager 1 is 1.3±0.1 and the mean ratio Cassini-to-Voyager 2 is 2.4±0.1.

**Fig. 8:** A comparison of the relative brightness of the F ring during six time periods. Each period consists of all observations taken during one calendar year. Fit curves are constrained to the same shape as the complete Cassini fit curve. Ratios of each year to the mean are 2004: 1.05, 2005: 0.93, 2006: 1.00, 2007: 0.99, 2008: 1.01, 2009: 0.98.

It may seem unfair for us to have removed the effects of the 2006 feature when looking for brightness changes. However, such a major feature was not present in the Voyager 1 images and was unlikely to have caused the 1.8 times brightness difference between Voyager 1 and Voyager 2. For completeness, we also performed the same per-year analysis on the Cassini images without adjusting for the effects of this feature. In this case, the mean of the measurements was 43% higher in 2007 and 6% higher in 2008. Even at its peak, the change produced was small compared to the change from Voyager 1 to Voyager 2. Thus, the longer-term changes in brightness are more likely due to global changes to the ring rather than to localized, transient features.

## 4. OCCULTATIONS

Stellar occultations can be used to measure the optical depth of the F ring directly. As in photometry, we calculate a single number that encompasses the total optical depth of the ring, the "normal equivalent depth" $D$:

$$D = \int \tau(a) da \qquad (4)$$

where $\tau(a)$ is the optical depth profile, normalized to be perpendicular to the ring plane (French et al., 1991).

The photopolarimeter (PPS) instrument on Voyager 2 (Lillie et al., 1977) produced a profile of the F ring on August 25, 1981 by observing the bright ultraviolet star δ Sco through a 0.264 µm filter as it was occulted by Saturn's rings (Lane et al., 1982). Observations were made every 10 msec resulting in a radial resolution of ~100 m. Analogous occultation profiles are available from Cassini's Visual and Infrared Mapping Spectrometer (VIMS) (Brown et al., 2004; Nicholson et al., 2007; Hedman et al., 2011) and Ultraviolet Imaging Spectrograph (UVIS) instruments (Esposito et al., 2004, 2008; Meinke et al., 2007, 2009, 2010; Colwell et al., 2010; Albers et al., 2012). We use the VIMS occultation data in this study.

While VIMS normally produces spatially-resolved spectra of planetary targets, it is also able to work in an occultation mode by disabling the imaging capabilities and taking IR spectra (0.85 μm – 5 μm with a resolution of 0.016 μm) from a single pixel targeted at a star. Each spectrum is then smoothed by co-adding into 0.13 μm bins. For this study, 30 occultations were chosen from those taken of the F ring by VIMS (Hedman et al., 2011). These were selected because they covered the complete radial range from the center of the A-ring/F-ring gap (Roche Division) to the same distance on the outside of the F ring, the same area covered by our ISS measurements, and had clean measurements without dropped or corrupted data. The data were not calibrated, but a mean instrument thermal background was subtracted and the data were normalized to an optical depth of zero on either side of the F ring (138,000–139,000 km and 141,000–142,000 km). The co-added spectral bin centered on 2.92 μm was used because scattered ring background is at a minimum at this wavelength. The details of the 30 occultations are listed in Table 1.

The F ring does not have a well-defined edge, and the optical depth is not precisely zero in our data set even in the area between the A ring and the F ring or outside of the F ring. To compare the PPS and VIMS occultation data properly, we recalibrated the PPS data using a technique as close as possible to that used for VIMS. This had the effect of slightly reducing the optical depths of the F ring core compared with those produced using the calibration method of Esposito et al. (1983).

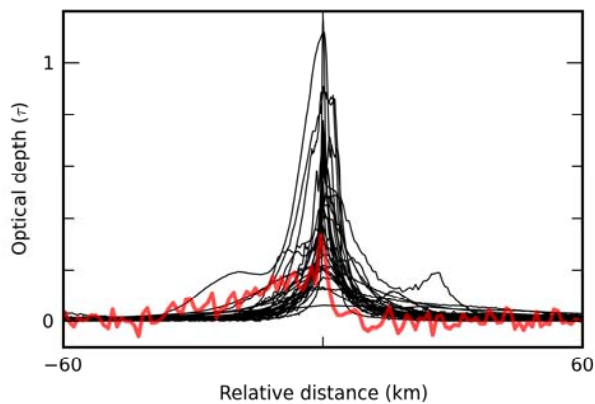

**Fig. 9: The 30 VIMS occultation profiles from Table 1 overlaid with the single Voyager 2 PPS occultation profile (red). Each profile is centered on the brightest part of the F ring core and limited to a 120 km width to better show the core detail.**

As areas outside the F ring have small, but non-zero, optical depths, computations of equivalent depth are sensitive to the distance limits of integration. For this paper, we used a 600 km width centered on the F ring core (defined as the location of the highest optical depth) for the VIMS data and a 120 km width for the PPS data. The 30 VIMS occultations and the single PPS occultation are shown in Fig. 9. The resultant equivalent depth for PPS is 4.57±0.12 km (where the error bar is due to uncertainties in the measured optical depths), consistent with the value of 4.33±0.13 km found by Showalter et al. (1992). The equivalent depths for VIMS are shown in Table 1. The mean equivalent depth is 9.99±2.65 km (where the error bar is the 1 σ scatter among the 30 occultations), approximately 2.2 times larger than the PPS value. This is consistent with the equivalent depth of 10±4 km found by Scharringhausen and Nicholson (2011) using a model of an inclined F ring derived from Hubble Space Telescope observations taken during the ring plane crossings of 1995.

| Star | Rev | I/E | Date | Ring opening angle (°) | Longitude (°) | Equivalent depth (km) | 90% width (km) |
|---|---|---|---|---|---|---|---|
| o Cet | 8 | I | 2005 MAY 24 05:00 | 3.45 | 171.54 | 8.40 | 358.4 |
| o Cet | 8 | E | 2005 MAY 24 08:04 | 3.45 | 29.98 | 8.29 | 398.7 |
| o Cet | 9 | I | 2005 JUN 11 08:06 | 3.45 | 51.38 | 7.27 | 393.3 |
| o Cet | 9 | E | 2005 JUN 11 10:25 | 3.45 | 306.88 | 7.36 | 405.5 |
| o Cet | 10 | I | 2005 JUN 29 12:39 | 3.45 | 258.25 | 6.93 | 421.1 |
| o Cet | 10 | E | 2005 JUN 29 14:24 | 3.45 | 181.94 | 7.87 | 378.8 |
| o Cet | 12 | E | 2005 AUG 05 01:57 | 3.45 | 207.65 | 11.27 | 349.7 |
| $\delta$ Vir | 29 | E | 2006 SEP 25 22:47 | 2.38 | 343.86 | 8.05 | 404.9 |
| $\delta$ Sco | 29 | I | 2006 SEP 26 06:35 | 32.16 | 264.36 | 8.69 | 329.2 |
| R Hya | 36 | I | 2007 JAN 01 16:27 | 29.40 | 92.19 | 6.98 | 427.2 |
| $\delta$ Sco | 55 | E | 2008 JAN 03 11:23 | 32.16 | 321.10 | 13.41 | 295.1 |
| R Leo | 60 | E | 2008 MAR 03 16:46 | 9.55 | 275.25 | 11.68 | 314.3 |
| R Leo | 61 | I | 2008 MAR 14 06:57 | 9.55 | 185.44 | 11.04 | 360.2 |
| R Leo | 61 | E | 2008 MAR 14 08:30 | 9.55 | 191.97 | 10.35 | 359.2 |
| R Leo | 63 | I | 2008 APR 03 12:33 | 9.55 | 277.49 | 8.51 | 378.2 |
| R Leo | 63 | E | 2008 APR 03 14:35 | 9.55 | 295.68 | 9.18 | 395.6 |
| $\gamma$ Cru | 73 | I | 2008 JUN 22 14:39 | 62.35 | 218.02 | 9.14 | 292.9 |
| CW Leo | 74 | I | 2008 JUL 02 00:16 | 11.38 | 33.88 | 10.00 | 402.3 |
| R Leo | 75 | E | 2008 JUL 09 06:59 | 9.55 | 193.27 | 12.29 | 360.8 |
| R Leo | 77 | I | 2008 JUL 23 06:22 | 9.55 | 261.15 | 9.07 | 383.9 |
| R Leo | 77 | E | 2008 JUL 23 09:03 | 9.55 | 272.76 | 7.37 | 434.2 |
| RS Cnc | 80 | I | 2008 AUG 13 01:14 | 29.96 | 23.55 | 12.82 | 347.3 |
| RS Cnc | 80 | E | 2008 AUG 13 08:18 | 29.96 | 323.76 | 9.00 | 397.3 |
| RS Cnc | 85 | I | 2008 SEP 18 21:39 | 29.96 | 178.61 | 10.77 | 391.0 |
| RS Cnc | 87 | I | 2008 OCT 03 15:26 | 29.96 | 240.33 | 15.94 | 286.6 |
| RS Cnc | 87 | E | 2008 OCT 03 22:20 | 29.96 | 180.00 | 9.47 | 437.3 |
| R Leo | 87 | E | 2008 OCT 04 20:23 | 9.55 | 337.24 | 10.22 | 475.0 |
| RS Cnc | 92 | I | 2008 NOV 10 00:42 | 29.96 | 99.99 | 9.62 | 384.6 |
| $\alpha$ Sco | 115 | I | 2009 JUL 27 22:14 | 32.16 | 149.32 | 18.68 | 199.1 |
| Mean | | | | | | 9.99±2.65 | 371.1±54.7 |

**Table 1: Selected Cassini VIMS occultations of the F ring evaluated over a 600 km distance centered on the F ring core. Rev = Cassini orbit number; I/E = ingress/egress (direction of star travel across the rings); Longitude = longitude in the same rotating reference frame used in Table 2.**

One might suspect a bias to our comparison because the wavelengths of the PPS and VIMS measurements differ, but the F ring's opacity has not shown a strong wavelength dependence between the UV and the near IR (Hedman et al., 2011; Albers et al., 2012). Furthermore, one would expect VIMS measurements to be reduced relative to PPS because small dust grains become generally less detectable as the wavelength increases.

Note that the VIMS equivalent depths discussed here are somewhat larger than those found by Hedman et al. (2011) using the same data, because they integrated a smaller radial slice of the F ring, the 150 km immediately surrounding the core, whereas we use the entire width of the F ring. This illustrates another bias that might arise from the difference in radial extents of the PPS and VIMS integrations. In the PPS data set, nothing is detectable outside the 120 km limits, so an additional integration would just incorporate more noise into the measurement. However, we can use the VIMS profiles to estimate this potential bias by deriving the mean ratio of 120-km integrals to 600-km integrals in the VIMS profiles. That ratio is 0.63, implying that the PPS equivalent depth might have been ~7.25 km if the signal-to-noise of that data set had been higher. This value is close to the 1 σ error on our VIMS measurements, but our VIMS error is based on statistical scatter. Thus the adjusted PPS measurement is less than ~85% of the ones from Cassini and the mean of the Cassini measurements remains consistently ~40% higher than the one from Voyager.

The remaining potential bias in this measurement is that the PPS profile, as a single measurement, may have sampled an atypical point in the F ring. However, Showalter et al. (1992) used a model of the F ring based on their photometric study to determine that the PPS occultation likely occurred at an "average" longitude and that the mean equivalent depth of the F ring was 5.0±0.3 km vs. 4.33±0.13 km measured by PPS. Furthermore, studies of F ring intensity show that the ring has bright clumps relative to its baseline, but rarely shows localized decreases in material. This implies that, if the PPS were an atypical measurement, it would be artificially high, not low. Thus, this possibility cannot account for the observed increase in equivalent depth from Voyager 2 to Cassini.

## 5. RING WIDTH

For this paper, we define the width of the F ring to be the smallest portion of a radial slice that contains 90% of the intensity of the total equivalent width; this slice need not be centered on the core, but certainly includes the core. We have found this value for each of our Cassini NAC images using the same radial profiles previously used to compute the equivalent widths. For each movie, we averaged these widths to produce a mean width for the entire F ring. The results are presented in Table 2. They show a mean width over all images of 580±70 km, where the error is the standard deviation of the total sample set. Adjustments for emission angle, incidence angle, and optical depth are irrelevant because they change only the absolute $I/F$, not the relative $I/F$ measurements within a single radial slice. We have experimented with other fractional widths, but all values from 50% to 95% show qualitatively the same behavior.

We also computed widths for 59 radial slices of resolved Voyager 1 images, yielding a mean width of 200±40 km, similar to the widths of 100–120 km measured by Smith et al. (1981). When compared to the Voyager 1 images, the Cassini mean ring width is nearly three times as large.

It is also enlightening to examine the variation of ring width with phase angle (Fig. 10). As can be seen, there is a distinct increase in width with increasing phase angle. This implies that the regions of the F ring outside of the central core have a higher fraction of (predominantly forward-scattering) dust than the core does. Although insufficient data are available from Voyager for a full analysis, the Voyager 1 data also show a similar trend with a mean width of 200±41 km at low phase (0–60°) and 247±9 km at mid phase (60–120°). The dependence on phase for low-to-mid phase angles, combined with the Cassini width measurements, implies that the non-core regions have consisted of micron-sized dust over 30 years. A full analysis of the particle size distribution using both ISS and VIMS data is planned for a future paper.

Finally, we can analyze the width of the F ring using the VIMS occultations. We define the width similarly, being 90% of the total equivalent depth. These data are shown in Table 1. The mean occultation width of 370±60 km is significantly smaller than the mean photometric width of 580±70 km. This suggests that the core of the F ring is sufficiently dense that its optical depth prevents all of the particles from being seen in reflected sunlight, further justifying the need for our new photometric model.

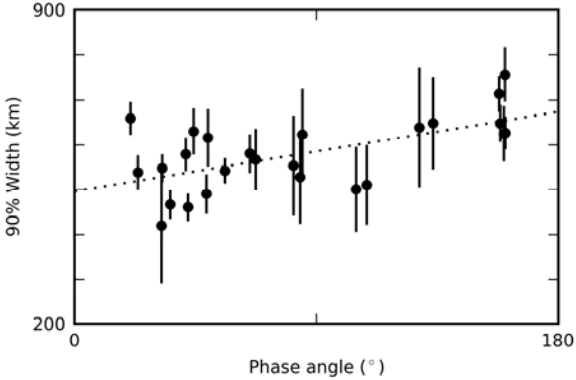 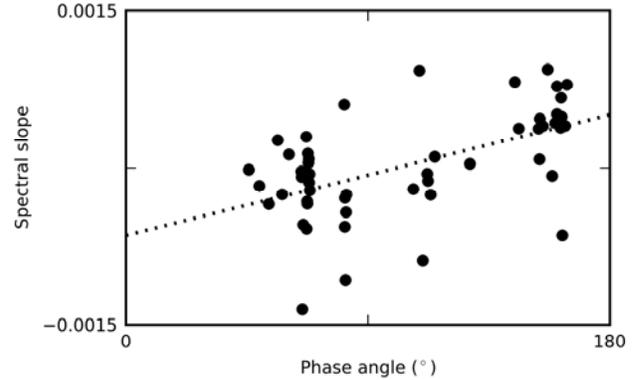

**Fig. 10:** The 90% width of the F ring versus phase angle for the Cassini NAC images. The line is the linear best fit and shows a distinct dependence of width on phase angle.

**Fig. 11:** The color of the F ring. Each point represents three Cassini NAC images taken through the RED, GRN, and BL1 filters. The line shows the best linear fit to the data and illustrates the color's dependence on phase.

## 6. INSTRUMENTAL CONSIDERATIONS

*6.1. Cross-calibration of Cassini and Voyager images*

Because we are comparing photometry from Cassini and Voyager images, it is important to verify that they have consistent cross-calibration. We selected a broad, featureless and (presumably) invariant region of the A ring, just interior to the Keeler gap, for this purpose. We identified nine pairs of Cassini and Voyager clear-filter images of the region (three from Voyager 1 and six from Voyager 2) that matched one another closely in phase and emission angles. For each image, a measurement was taken by averaging *I/F* over a large rectangular area of the selected region. We compensated for differences in solar incidence angle *i* by dividing each measurement by cos(*i*), which adjusts for the relative "spread" of sunlight over the ring's surface. The mean ratio of each Cassini measurement to its Voyager counterpart was 0.95±0.23 (1 σ). Several pairs of images with similar incidence angles (within 0.6°) are also available. In these cases we were able to match the emission angle very closely, but were unable to match the phase angle to better than ~10°. In these cases we performed no geometric correction and continued to find similar measurements between Voyager and Cassini. In those cases where there was a significant difference, the Voyager 1 observation was always brighter. This shows that the absolute calibration of the Cassini and Voyager image processing pipelines are statistically consistent. The remaining uncertainty, which is biased in a direction opposite our findings, is much smaller than the magnitude of intrinsic variations in the F ring.

Another potential source of calibration error is color. Voyager's vidicon-based cameras are blind to red wavelengths, whereas the passbands of Cassini's CCD-based cameras extend well into the near infrared. Thus, if the F ring is not neutral in color, this could bias any comparison between Voyager and Cassini results. To measure the ring color, we chose 51 sets of Cassini F ring images covering a wide range of phase and emission angles, each taken in quick succession through the BL1 ($\lambda_{eff}$ = 455 nm), GRN ($\lambda_{eff}$ = 569 nm), and RED ($\lambda_{eff}$ = 649 nm) filters. Equivalent width $W$ was calculated for the three images in each set, and a least-squares linear fit was performed to determine the spectral slope. The fit was then evaluated at the RED and BL1 effective wavelengths (Fig. 11). The average ratio $W_{RED}/W_{BL1}$ was 1.01±0.10, which is very nearly neutral. The ratios show a detectable dependence on the phase angle, with the F ring slightly bluer at low phase and slightly redder at high phase. However, this dependence is weak and should not significantly alter our results.

*6.2. Dynamic range*

The Voyager camera system used 8-bit digitization and a linear compression scheme, while the Cassini cameras use 12-bit digitization and (typically) logarithmic compression. This allows the Cassini cameras to have a much wider dynamic range. Although the absolute calibration appears to be consistent for bright objects, it is possible that the Voyager cameras are simply less sensitive to dim objects. This would account for both the lower mean brightness and the smaller radial widths. However, the Voyager 1 and Voyager 2 cameras were essentially identical and yet we find dramatic differences in mean brightness between those two missions. Such a difference could not be explained by instrumentation differences.

The spatial resolution of the cameras could also yield some differences. The point spread function (PSF) of the Voyager NACs cover a diameter of ~2 pixels (Karkoschka, 2001), while the Cassini PSF covers approximately a single pixel (Porco et al., 2004). Because of the lower resolution of some of the Voyager images, only a few pixels separate the A ring from the F ring. This could result in the light from the A ring bleeding over into the gap between the two rings and, by artificially raising the background level, reducing the inferred equivalent width. Although this problem could affect a number of the Voyager 1 images, only a few Voyager 2 images have marginal resolution. Because the Voyager 1 images are systematically brighter than the Voyager 2 images, exactly the opposite of the behavior we might expect, we conclude that this does not play a significant role in our photometry.

We also tested the possibility that effects such as these could have biased our measurements of the ring's width. We processed the Cassini radial profiles to make them as close as possible to the quality of Voyager images. This involved down-sampling the resolution to 50 km per pixel and convolving the profiles with the PSF of the Voyager camera, found by Karkoschka (2001) to be well-represented by a Gaussian with a full width at half maximum of two pixels. The mean widths of the F ring before and after this modification differed by at most 10%.

A similar issue arises with the stellar occultation data. The Voyager PPS instrument is insensitive to optical depths below ~0.01 (Showalter et al., 1992). It is easy to believe that the removal of the low-optical-depth non-core regions of the F ring could reduce the Cassini-measured equivalent depths to the Voyager level. To address this concern, we re-computed the Cassini equivalent depths removing any optical depth less than 0.01. The new mean equivalent depth is 7.9±2.8 km, still nearly twice the ~4.6 km measured by Voyager 2. Thus, we do not believe that the lesser sensitivity of the PPS data can explain the ring's apparent temporal change.

# 7. DISCUSSION

## 7.1. The role of Prometheus

Whenever periodic changes in the F ring are seen, it is tempting to relate them to interactions between the ring and its inner shepherd moon, Prometheus. The ring and Prometheus precess at slightly different rates, and the time from one pericenter alignment to the next is approximately 17 years (Chavez, 2009). Based on recent orbital measurements (Bosh et al., 2002; Jacobson et al., 2008), the F ring and Prometheus reached pericenter anti-alignment in early 1975, mid 1992, and late 2009. Near these times Prometheus's orbit actually crosses the inner components of the F ring, although the precise distance of closest approach is difficult to predict because of the variability of Prometheus's eccentricity (Borderies and Goldreich, 1983).

That a nearby satellite can influence the formation and destruction of clumps and the production of more complicated structures such as "braids" has been well established by both theory and observation (Lissauer and Peale, 1986; Hänninen, 1993; Lewis and Stewart, 2005; Murray et al., 2005, 2008; Winter et al., 2007; Beurle et al., 2010; Meinke et al., 2010; Esposito et al., 2012), and thus Prometheus may have influenced the changes in brightness and width of the F ring over the past 30 years. During the Voyager flybys in 1980–1981, the distance of closest approach increased from ~475 km to ~550 km, and during the five years of the Cassini mission used in this study the distance decreased from ~450 km to ~200 km. The increase in brightness of Voyager 1 over Voyager 2 could be explained by the closer approaches of Prometheus. However, we have found no systematic change in the brightness of the F ring during the five years of the Cassini tour, a period during which the proximity of Prometheus changed more dramatically. The data thus appear to contradict the hypothesis that Prometheus is the primary cause of long-term changes in the F ring.

## 7.2. The F ring environment

The F ring exists in a unique environment. Its position ~140,000 km from Saturn's center places it slightly outside of Saturn's classical Roche limit and yet firmly within the Roche zone. Here, Saturn's tides are not strong enough to prevent gravitational accretion, but are strong enough to slow accretion. In particular, numerical simulations and theoretical analyses (Canup and Esposito, 1995, 1997; Throop and Esposito, 1998; Barbara and Esposito, 2002; Karjalainen and Salo, 2004; Karjalainen, 2007) have shown that such tidally-modified accretion tends to result in a bimodal particle size distribution, with a small number of larger bodies coexisting with a large number of much smaller ring particles and dust. This is because, in this environment, only particles with sufficiently different masses can remain gravitationally bound, while similar-mass bodies overflow their mutual Hill sphere. In the environment of the F ring, the required mass ratio ranges from a few to ~100, depending on assumptions about the coefficient of restitution, surface friction, and particle velocity distribution. Interestingly, for certain reasonable assumptions, the core of the F ring exists within a small transition zone region (Karjalainen, 2007); any closer to Saturn results in net destruction of aggregates, while any farther results in net accretion. Thus the F ring may be precariously balanced between forces of destruction and accretion and can be easily influenced by other processes.

The continued existence of fine dust in the F ring is troublesome, as Poynting-Robertson drag and other forces should cause tiny particles to spiral inward to the A ring in far less than $10^6$ years. Micron-sized dust should be destroyed by micrometeoroid impacts or erosion by high-energy magnetospheric particles in only tens to hundreds of thousands of years (Burns et al., 1984). The existence of "parent" aggregate bodies inside the ring has long been postulated to solve this problem (Smith et al., 1982; Lissauer and Peale, 1986; Showalter et al., 1992; Murray et al., 1997). These same parent bodies probably supply the core of mass that enables the ring to maintain its eccentricity and precess as a single unit in spite of its radial width (Murray et al., 1997; Bosh et al., 2002). A small, opaque, sharp-edged feature that appeared in one UVIS and VIMS occultation may represent the first direct detection of one of these bodies (Nicholson et al., 2007; Esposito et al., 2008). Esposito et al. (2008) used the UVIS occultation data to estimate that there are between 1,500 and 140,000 moonlets ~600 m in size in the F ring's core.

In addition to parent bodies, additional bodies orbit nearby. These were first inferred from unexplained drops in trapped magnetospheric electrons within a 2000-km-wide band around the F ring (Cuzzi and Burns, 1988). Kolvoord et al. (1990) found periodic signals not associated with Prometheus in longitudinal profiles from Voyager images and suggested that these provided evidence for small, undiscovered satellites near the F ring as well, although such signals have not been reported in Cassini data. Since then, small moonlets or elongated clumps have been found on neighboring orbits interior to the F ring (McGhee et al., 2001) or that cross the F ring's core (Porco et al., 2005; Spitale et al., 2006).

The F ring's prevalent dust comprises the (very) loosely bound regolith of these parent bodies. The ring's reflectivity is determined by the cross-sectional area of this material, so a release of dust particles can produce a marked increase in ring brightness, even while the mass of the ring overall is conserved. As the width dependence on the phase angle shows, the regions outside the core are composed primarily of these smaller particles. This is consistent with measurements by Hedman et al. (2011), who used the spectral features of the F ring during VIMS stellar occultations to determine that larger particles are concentrated near the core, and by Showalter et al. (1992), who came to a similar conclusion by comparing stellar occultations at different wavelengths using the Voyager PPS and Radio Science Subsystem (RSS) instruments. Thus the brightening of the F ring is consistent with a core of larger aggregates and the periodic release of regolith to form clumps.

Prometheus corrals the F ring material into groupings with a characteristic spacing of 3.2°, but impacts and collisions are likely responsible for the less regular and far brighter clumps. Showalter (1998; 2004) tracked these features through the Voyager encounters, detecting three distinct "burst" events within a time span of a few months. However, he found that most clumps persisted for the duration of one Voyager flyby, but nothing lasted for the nine months between the two Voyager encounters. This contrasts with the large event of late 2006, which was still detectable two years later, and supports the conclusion that this was the result of an unusually large impact. During Earth's crossings through the Saturn ring plane in 1995, the F ring was continuing to show clumps (Bosh and Rivkin, 1996; Nicholson et al., 1996; McGhee et al., 2001).

Based on the inferred high impact velocities required to create the three sudden bursts, Showalter (1998) ascribed them to meteoroid impacts. Barbara and Esposito (2002) argued that collisions between the parent bodies were responsible, despite the very slow speeds with which they occur. Neither set of authors discussed the possibility of bodies on neighboring orbits, which could collide into the ring's core at intermediate velocities. In reality, it is this neighboring population that appears to be responsible for the large clumps. Murray et al. (2008) showed that the event of late 2006 was triggered by the impacts of one such body, S/2004 S6, into the core of the ring. They also found that other "fans" are likely caused by impacts from bodies on orbits slightly different from that of the ring.

*7.3. Ring evolution*

To summarize our results, the F ring has increased in overall brightness, optical depth, and radial width from the Voyager era to the present. These changes are generally factors of two to three. The F ring of today contains substantially more dust than it did 30 years ago, although its overall mass may well remain unchanged. Unexpectedly, we also note a significant change in the ring's overall brightness between the Voyager 1 and Voyager 2 encounters, which were separated by only nine months. In spite of these changes, however, the ring's properties have been nearly stable for the last five years.

We have also noted that a very large impact occurred in late 2006. This appears to have been associated with dust clump S/2004 S6, which apparently followed a different orbit than the F ring and slowly, due to relative apsidal precession and relative nodal regression, found itself on a collision course with the ring's core. Beyond the observations summarized above, one of the most puzzling aspects of today's F ring is what we do not see. Aside from the 2006 event, the ring has shown relatively few impacts in the last five years. This is apparent in the "before" profile (dotted) of Fig. 4c, where the ring is nearly uniform in longitude. By contrast, Voyager and Hubble images always showed two to three features each locally two to three times as bright as the rest of the ring (Nicholson et al., 1996; Showalter, 1998). In the Cassini data, the most recent similar event was one that resulted in a kinematic spiral, as described by Charnoz et al. (2005), which was already spreading and dissipating at the time of Cassini's arrival in 2004.

A decrease in the number of ring impacts suggests a decrease in the number of impactors. We hypothesize that the family of potential impactors on nearby orbits, such as S/2004 S6, may be substantially smaller today than it was in the early 1980s. Perhaps the outcome of each impact is the appearance of one transient bright "burst" in the ring, but one fewer clump on a nearby orbit; the F ring is gradually absorbing the bodies in its vicinity. If this hypothesis is correct, then one might suppose that the ring is remarkably young. To play the process backward beyond 1980, the family of nearby bodies must have been substantially larger 60 or 90 years ago. Eventually, that population becomes larger than the ring itself, at which point we do not have a ring so much as a swarm of debris, perhaps from the outcome of a large collisional breakup.

A ring with a lifetime measured in decades or centuries would certainly challenge our assumptions about the rate of evolution of Saturn's small inner moons. Perhaps more plausibly, a mechanism exists whereby the ring itself can eject bodies like S/2004 S6, which follow nearby trajectories and inevitably re-impact the ring years or decades later. If so, then the F ring region is stable overall, but it undergoes variations on time scales of decades.

The growth in ring width from the 1980s to the present may hold a clue to this process. It has long been noted that the F ring precesses as a whole. This is very hard to understand in light of its large radial width, because something must act to suppress the differential apsidal precession and nodal regression that one would expect. The most plausible explanation is that the F ring's core holds enough mass (at least $10^{-9}$ Saturn masses according to Murray et al. (1997)) to keep its outliers in lockstep. If so, then the ring's width is probably defined by the limits over which this process can operate; if it were to grow wider, then the most distant material would break loose from the ring, precess freely, and perhaps coalesce into new bodies like S/2004 S6. In this view, the ring spawns its own impactors.

As noted above, we cannot connect any observable trend in the ring's properties to the 17-year cycle of its mutual configuration with Prometheus. Nevertheless, the images clearly show that Prometheus stirs up the ring during the periods of apsidal anti-alignment. Because this interaction follows a similar, decadal time scale, perhaps we should not be too hasty in dismissing its role. One can imagine a scenario where Prometheus stirs up the ring, the ring grows broader, material breaks loose, and only later re-impacts the ring. In such a process, variations in the rate of the observed, large impacts into the ring could fall substantially out of phase with the times when Prometheus is penetrating most deeply into the ring.

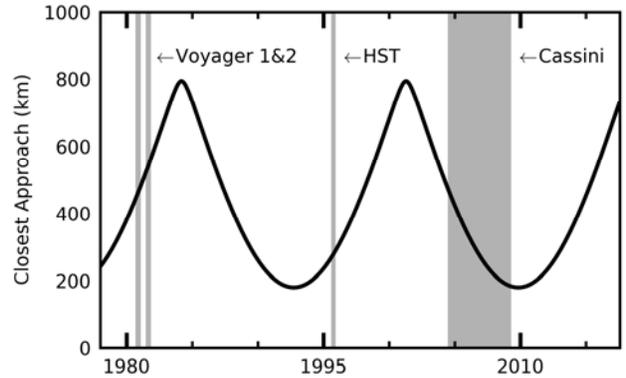

Fig. 12: The distance of closest approach between Prometheus and the F ring core. The observations from Voyager 1&2, HST, and Cassini (dates used in this work) are highlighted.

We close by offering the general outline of a model that can explain all of our seemingly contradictory measurements. During 1980 and 1981, the ring was "cooling off" as the effects of the most recent Prometheus passages were fading (Fig. 12). As a result, it was narrow and faint, and the ring became measurably fainter during the nine months between the two Voyager encounters. However, at the same time, the material that had broken free during the last apsidal anti-alignment was still coalescing and re-impacting the ring. These impacts were frequent and dramatic, producing the three observed bursts and the numerous other clumps so prevalent at the time. Observations of the ring during the ring plane crossings of 1995 (another "cooling off" period) were similar, also showing multiple large clumps as well as nearby moonlets (McGhee et al., 2001).

The Cassini era has been very different, with Prometheus moving deeper into the ring rather than receding. Even by the time of Cassini's arrival in 2004, Prometheus was already entering deep enough into the ring to keep it "hot." Throughout the Cassini tour, the ring has been at its maximum brightness and width; it cannot grow wider without shedding material. On the other hand, because so much time has elapsed since the prior anti-alignment, most of the material that escaped at that time has already re-impacted the ring's core. Only a small amount of material remains on nearby orbits, so impacts have been rare. Nevertheless, the occasional large impact still occurs, as one did in 2006.

This model certainly requires further data analysis and dynamical modeling to determine if it is both plausible and consistent with the data. However, it makes predictions that can be tested in the Cassini Solstice Mission, set to continue into 2017. One would expect the ring to remain stable in width and brightness for the time being, but eventually to cool off as Prometheus backs away. After a few years, it will begin to fade in brightness and narrow in width. At the same time, the number of identifiable impacts into the ring will slowly increase. By 2016, the ring and Prometheus will be in an orbital configuration similar to where they were at the Voyager encounters, and the ring's properties will once again be similar to what we observed so long ago.

APPENDIX

Here we describe our method for modeling shadowing and obscuration. We define a new quantity, $U$, which is the hypothetical value of $I/F$ that would be measured normal to the ring plane if every ring particle were fully illuminated and unobscured. In all cases, $U \geq \mu I/F$, because shadowing and obscuration always reduce the reflected or transmitted sunlight. If a ring has sufficiently low optical depth that shadowing and obscuration can be ignored, $U$ reduces to our previous $\mu I/F$:

$$U = \frac{I}{F}\bigg|_{\perp, \tau \ll 1} = \mu \frac{I}{F}\bigg|_{\tau \ll 1} = \frac{\varpi_0 P(\alpha) \tau}{4},$$  (A.1)

where $\varpi_0$ is the single-scattering albedo and $P(\alpha)$ is the phase function in terms of the phase angle $\alpha$.

Chandrasekhar (1960) (see also Hansen and Travis, 1974; Dones et al., 1993) derived an analytic expression for reflected intensity in the presence of single-scattering:

$$\frac{I}{F} = \mu_0 R(\tau, \mu, \mu_0, \alpha),$$  (A.2)

where

$$R(\tau, \mu, \mu_0, \alpha) = \frac{\varpi_0 P(\alpha)(1 - \exp(-\tau(1/\mu + 1/\mu_0)))}{4(\mu + \mu_0)}.$$  (A.3)

For transmitted intensity, applicable to measurements on the unlit face of the ring plane,

$$\frac{I}{F} = \mu_0 T(\tau, \mu, \mu_0, \alpha),$$  (A.4)

where

$$T(\tau, \mu, \mu_0, \alpha) = \frac{\varpi_0 P(\alpha)(\exp(-\tau/\mu) - \exp(-\tau/\mu_0))}{4(\mu - \mu_0)}.$$  (A.5)

Here $\mu_0 \equiv |\cos(i)|$. In the low-$\tau$ limit, the formulas reduce via Taylor expansion to:

$$R_{\tau \ll 1}(\tau, \mu, \mu_0, \alpha) = T_{\tau \ll 1}(\tau, \mu, \mu_0, \alpha) = \frac{\varpi_0 P(\alpha) \tau}{4 \mu \mu_0}$$  (A.6)

$$\frac{I}{F} = \mu_0 \frac{\varpi_0 P(\alpha) \tau}{4 \mu \mu_0} = \frac{\varpi_0 P(\alpha) \tau}{4 \mu}$$  (A.7)

which is consistent with Eq. A.1. As $\tau$ increases, $U$ and $I/F_\perp$ diverge, with $U$ increasing linearly but $I/F_\perp$ trending toward smaller values. To convert from $I/F_\perp$, which we know from measurements and viewing geometry, to $U$, we solve for the ratio:

$$z = \frac{U}{I/F_\perp}. \tag{A.8}$$

It is evident that $z \to 1$ as $\tau \to 0$. Because the functions $R$ and $T$ compute the fraction of light reflected or transmitted in the presence of shadowing and obscuration and $R_{\tau \ll 1} = T_{\tau \ll 1}$ computes the fraction of light reflected without shadowing or obscuration, we have:

$$z_R(\tau, \mu, \mu_0) = \frac{R_{\tau \ll 1}(\tau, \mu, \mu_0, \alpha)}{R(\tau, \mu, \mu_0, \alpha)} = \frac{\tau(\mu + \mu_0)}{\mu\mu_0(1 - \exp(-\tau(1/\mu + 1/\mu_0)))} \tag{A.9}$$

and

$$z_T(\tau, \mu, \mu_0) = \frac{T_{\tau \ll 1}(\tau, \mu, \mu_0, \alpha)}{T(\tau, \mu, \mu_0, \alpha)} = \frac{\tau(\mu - \mu_0)}{\mu\mu_0(\exp(-\tau/\mu) - \exp(-\tau/\mu_0))}, \tag{A.10}$$

with $z = z_R$ or $z = z_T$ as appropriate. Note that $z$ no longer depends on $\alpha$ or $\varpi_0$.

It is now possible to compute $U$ given $\mu$, $\mu_0$, and $\tau$.

$$U = z\frac{I}{F_\perp} = z\mu\frac{I}{F}. \tag{A.11}$$

We can define a "$\tau$-corrected equivalent width" as:

$$W_\tau(\tau, \mu, \mu_0) = \int U(a)da = \int z(\tau(a), \mu, \mu_0)\mu\frac{I}{F}(a)da. \tag{A.12}$$

This quantity is, however, tied to the radial profile of optical depth $\tau(a)$, which is known to be highly variable from occultation data (Esposito et al., 2008; Hedman et al., 2011; Albers et al., 2012). Our approach is to introduce a single value $\tau_{equiv}$ to define the mean properties of the ring:

$$W_\tau(\tau, \mu, \mu_0) = z(\tau_{equiv}, \mu, \mu_0)\mu\int \frac{I}{F}(a)da, \tag{A.13}$$

where

$$z(\tau_{equiv}, \mu, \mu_0) = \int z(\tau(a), \mu, \mu_0)da. \tag{A.14}$$

ACKNOWLEDGMENTS

This work was funded by NASA's Cassini Data Analysis Program through Grants NNX07AJ76G and NNX09AE74G. This research project was initiated during the COSPAR Regional Workshop on Planetary Science, 23 July–3 August 2007, Montevideo, Uruguay.

| Date | Cassini Observation ID | First Image | Last Image | Number of images | Longitudinal resolution (°/image) | Longitude range (°) |
|---|---|---|---|---|---|---|
| 2004 JUN 20 | ISS_000RI_SATSRCHAP001_PRIME | N1466448701_1 | N1466504861_1 | 84 | 6.5 | 0-359 |
| 2004 NOV 15 | ISS_00ARI_SPKMOVPER001_PRIME | N1479201492_1 | N1479254052_1 | 73 | 4.8 | 276-269 |
| 2005 APR 13 | ISS_006RI_LPHRLFMOV001_PRIME | N1492052646_1 | N1492102189_1 | 1334 | 0.3 | 29-354 |
| 2005 MAY 01 | ISS_007RI_LPHRLFMOV001_PRIME | N1493613276_1 | N1493662416_1 | 247 | 1.2 | 138-101 |
| 2006 SEP 28 | ISS_029RF_FMOVIE001_VIMS | N1538168640_1 | N1538218132_1 | 93 | 3.5 | 297-266 |
| 2006 OCT 31 | ISS_031RF_FMOVIE001_VIMS | N1541012989_1 | N1541062380_1 | 112 | 2.9 | 48-16 |
| 2006 NOV 12 | ISS_032RF_FMOVIE001_VIMS | N1542047155_1 | N1542096546_1 | 112 | 2.9 | 282-250 |
| 2006 NOV 25 | ISS_033RF_FMOVIE001_VIMS | N1543166702_1 | N1543216891_1 | 99 | 3.3 | 133-108 |
| 2006 DEC 23 | ISS_036RF_FMOVIE001_VIMS | N1545556618_1 | N1545613256_1 | 128 | 3.0 | 0-359 |
| 2007 JAN 05 | ISS_036RF_FMOVIE002_VIMS | N1546700688_5 | N1546748805_1 | 125 | 2.4 | 64-20 |
| 2007 FEB 10 | ISS_039RF_FMOVIE002_VIMS | N1549801218_1 | N1549851279_1 | 124 | 2.6 | 42-7 |
| 2007 FEB 27 | ISS_039RF_FMOVIE001_VIMS | N1551253524_1 | N1551310298_1 | 144 | 2.6 | 0-359 |
| 2007 MAR 17 | ISS_041RF_FMOVIE002_VIMS | N1552790437_1 | N1552850917_1 | 169 | 2.4 | 0-359 |
| 2007 MAR 31 | ISS_041RF_FMOVIE001_VIMS | N1554026927_1 | N1554072073_1 | 128 | 2.3 | 337-277 |
| 2007 MAY 05 | ISS_044RF_FMOVIE001_VIMS | N1557020880_1 | N1557086720_1 | 178 | 2.5 | 0-359 |
| 2007 OCT 18 | ISS_051RI_LPMRDFMOV001_PRIME | N1571435192_1 | N1571475337_1 | 258 | 1.3 | 100-8 |
| 2007 DEC 31 | ISS_055RF_FMOVIE001_VIMS | N1577809417_1 | N1577857957_1 | 149 | 2.2 | 313-275 |
| 2008 JAN 07 | ISS_055RI_LPMRDFMOV001_PRIME | N1578386361_1 | N1578440131_1 | 191 | 1.8 | 302-302 |
| 2008 JAN 23 | ISS_057RF_FMOVIE001_VIMS | N1579790806_1 | N1579838190_2 | 132 | 2.3 | 268-224 |
| 2008 MAY 15 | ISS_068RF_FMOVIE001_VIMS | N1589589182_1 | N1589641908_1 | 40 | 8.5 | 107-92 |
| 2008 JUL 05 | ISS_075RF_FMOVIE002_VIMS | N1593913221_1 | N1593969867_1 | 110 | 3.5 | 0-359 |
| 2008 AUG 30 | ISS_083RI_FMOVIE109_VIMS | N1598806665_1 | N1598853071_1 | 222 | 1.4 | 317-267 |
| 2008 SEP 30 | ISS_087RF_FMOVIE003_PRIME | N1601485634_1 | N1601526770_1 | 171 | 1.6 | 117-33 |
| 2008 OCT 14 | ISS_089RF_FMOVIE003_PRIME | N1602717403_1 | N1602760410_1 | 176 | 1.6 | 86-15 |
| 2009 JAN 11 | ISS_100RF_FMOVIE003_PRIME | N1610364098_1 | N1610404395_1 | 211 | 1.3 | 288-190 |

| Date | Cassini Observation ID | Radial resolution (km/pixel) | Phase angle (°) | Emission angle (°) | Incidence angle (°) | Equivalent width (km) | 90% width (km) |
|---|---|---|---|---|---|---|---|
| 2004 JUN 20 | ISS_000RI_SATSRCHAP001_PRIME | 40.3 | 67 | 74 | 65 | 2.6±0.8 | 567.3±68.6 |
| 2004 NOV 15 | ISS_00ARI_SPKMOVPER001_PRIME | 27.0 | 85 | 77 | 67 | 3.1±0.6 | 623.2±102.0 |
| 2005 APR 13 | ISS_006RI_LPHRLFMOV001_PRIME | 7.5 | 36 | 84 | 68 | 4.7±0.9 | 466.5±32.3 |
| 2005 MAY 01 | ISS_007RI_LPHRLFMOV001_PRIME | 7.4 | 32 | 69 | 68 | 1.5±0.3 | 419.5±128.9 |
| 2006 SEP 28 | ISS_029RF_FMOVIE001_VIMS | 9.4 | 160 | 122 | 74 | 12.6±2.7 | 756.6±59.9 |
| 2006 OCT 31 | ISS_031RF_FMOVIE001_VIMS | 9.4 | 158 | 126 | 75 | 10.3±1.4 | 648.2±41.3 |
| 2006 NOV 12 | ISS_032RF_FMOVIE001_VIMS | 9.3 | 158 | 126 | 75 | 9.9±1.8 | 714.0±40.1 |
| 2006 NOV 25 | ISS_033RF_FMOVIE001_VIMS | 10.0 | 160 | 121 | 75 | 12.9±1.7 | 626.9±36.8 |
| 2006 DEC 23 | ISS_036RF_FMOVIE001_VIMS | 11.8 | 160 | 124 | 76 | 13.6±5.3 | 624.9±61.9 |
| 2007 JAN 05 | ISS_036RF_FMOVIE002_VIMS | 10.2 | 133 | 144 | 76 | 2.9±2.2 | 648.5±103.4 |
| 2007 FEB 10 | ISS_039RF_FMOVIE002_VIMS | 10.3 | 128 | 148 | 76 | 2.7±1.7 | 639.1±134.0 |
| 2007 FEB 27 | ISS_039RF_FMOVIE001_VIMS | 10.2 | 105 | 145 | 76 | 1.7±1.0 | 500.6±95.1 |
| 2007 MAR 17 | ISS_041RF_FMOVIE002_VIMS | 10.8 | 109 | 144 | 77 | 1.8±1.0 | 510.7±89.2 |
| 2007 MAR 31 | ISS_041RF_FMOVIE001_VIMS | 11.9 | 84 | 129 | 77 | 2.1±0.9 | 528.2±104.7 |
| 2007 MAY 05 | ISS_044RF_FMOVIE001_VIMS | 12.4 | 82 | 119 | 77 | 2.4±0.9 | 553.5±111.8 |
| 2007 OCT 18 | ISS_051RI_LPMRDFMOV001_PRIME | 14.0 | 56 | 94 | 80 | 8.1±1.6 | 542.1±28.4 |
| 2007 DEC 31 | ISS_055RF_FMOVIE001_VIMS | 10.0 | 65 | 123 | 81 | 1.3±0.3 | 580.4±43.3 |
| 2008 JAN 07 | ISS_055RI_LPMRDFMOV001_PRIME | 9.4 | 21 | 101 | 81 | 3.2±0.5 | 659.2±36.8 |
| 2008 JAN 23 | ISS_057RF_FMOVIE001_VIMS | 10.3 | 44 | 118 | 81 | 1.3±0.3 | 630.2±51.8 |
| 2008 MAY 15 | ISS_068RF_FMOVIE001_VIMS | 6.5 | 50 | 130 | 83 | 0.9±0.1 | 616.0±64.6 |
| 2008 JUL 05 | ISS_075RF_FMOVIE002_VIMS | 6.9 | 33 | 116 | 84 | 1.2±0.2 | 547.8±31.2 |
| 2008 AUG 30 | ISS_083RI_FMOVIE109_VIMS | 7.2 | 24 | 107 | 85 | 1.9±0.6 | 538.4±38.4 |
| 2008 SEP 30 | ISS_087RF_FMOVIE003_PRIME | 5.8 | 49 | 130 | 85 | 0.9±0.2 | 490.2±44.3 |
| 2008 OCT 14 | ISS_089RF_FMOVIE003_PRIME | 6.4 | 42 | 122 | 85 | 1.0±0.2 | 460.3±30.9 |
| 2009 JAN 11 | ISS_100RF_FMOVIE003_PRIME | 6.3 | 41 | 128 | 87 | 0.8±0.1 | 578.9±37.6 |

**Table 2: Summary of Cassini NAC images used. All images are part of a movie with first and last image numbers as shown. Longitude is relative to a reference frame co-rotating with the F ring with a mean motion of 581.979°/day (Albers et al., 2012) with an origin at the intersection of the ascending node of Saturn's ring plane and Earth's equator at the epoch 1 JAN 2004 12:00 UTC. Longitude ranges starting at a higher number and going to a lower one wrap around past 360°. Emission and incidence angles are relative to the normal vector on the lit side of the ring plane.**